\documentclass[showpacs]{revtex4}

\usepackage{graphicx}
\usepackage{dcolumn}
\usepackage{amsmath}

\makeatletter
\def\btt#1{\texttt{\@backslashchar#1}}
\DeclareRobustCommand\bblash{\btt{\@backslashchar}}
\makeatother

\begin{document}
\DeclareGraphicsRule{.jpg}{eps}{.jpg.bb}{`jpeg2ps -r0 #1}

\title[]{Embedding Diagrams for the Reissner-Nordstr\"om spacetime}

\author{Aseem Paranjape}\thanks{Visiting student, IUCAA, Pune-411 007, INDIA}
\email{p_aseem83@rediffmail.com}
\affiliation{%
St.Xavier's College, Mahapalika Marg
Mumbai-400 001, INDIA}%

\author{Naresh Dadhich}
\email{nkd@iucaa.ernet.in}
\affiliation{%
Inter University Centre for Astronomy and Astrophysics,
Post Bag 4, Ganeshkhind,
Pune-411 007,INDIA}%

\date{\today}

\begin{abstract}
We consider embedding diagrams for the Reissner-Nordstr\"om
spacetime. We embed the $(r-t)$ and $(r-\phi)$ planes into 
$3$-Minkowski/Euclidean space and
discuss the relation between the diagrams and the corresponding curvature
scalar of the $2$-metrics.
\end{abstract}

\pacs{ 0420, 0420B, 0420J}

\maketitle

Embedding of a curved surface in the Euclidean space visually demonstrates
the ``curvedness'' or curvature of the surface. Embedding diagrams therefore
become excellent tools for visualizing the curvature of spacetime;i.e. the
gravitational field. The particle trajectories could easily be seen and they
would be analogues of the field lines in electromagnetic theory - they are
indeed field lines of gravitational field. Of the most interesting objects in
this context is black hole causing critical warping of space around it so
that no null ray can propagate out of it. The embedding diagram would make
this phenomenon transparent and visually illuminating.

In \cite{dm}, a departure is made, from the conventional embedding process,
in that the $(r-t)$ plane of spherically symmetric spacetimes is embedded into
$(2+1)$-Minkowski spacetime. Conventional embeddings are generally restricted 
to constant-time slices of the geometry and hence convey information only
 regarding the spatial curvature. Intuitively speaking the kinematic part of
the field, gravitational potential sits in $g_{tt} = 1 +2\Phi$, while it is
$g_{rr}$ which brings in the contribution of gravitational field energy
\cite{n0,n1,n2}. The latter is purely relativistic feature which is absent in
the Newtonian theory and is in fact the distinguishing characteristic of
general relativity (GR). When we embed the $t=const.$ slice, we are purely
considering the spatial curvature which hinges on the field energy while
$(r-t)$ plane would have both the gravitational potential as well as the field
energy included and hence would refer to the entire field. The contributions
from the both should act in unison and it can be demonstrated that the space
curvature does indeed guide free particle towards the centre \cite{n1} and
for this to happen it must be negative which would demand the field energy to
be negative \cite{n2}. Thus the {\it positive energy} condition for the field
energy density is that it be {\it negative}.

 The main aim of this note is for the Reissner-Nordstr\"om (RN) black hole to
relate the scalar curvature of the surface being embedded with the
character of the embedding diagram. The $(r-t)$ plane embedding
has been studied \cite{dm} and we follow the same method (which
was first employed in \cite{dm1}) for the spatial embedding. Of
course the embedding has to be carried out over patches for $r_{+}\le
r, \, r_{-}\le r\le r_{+}, \, Q^2/2M\le r\le r_{-}$ where $r_{\pm} =
M\pm\sqrt{M^2 - Q^2}$ are the two horizons. In the former case, it
cannot go as far in as $r_{-}$ and can even stop above $r_{+}$ if
$Q^2/M^2> 8/9$, while the spatial embedding can go down to $r = Q^2/2M$,
the hard core radius, covering all the three patches. Of
particular interest is the patch below $r_{-}$.

 We begin with RN spacetime metric described by
\begin{equation}
ds^2 = -\left(1 - \frac{2M}{r} + \frac{Q^2}{r^2} \right)dt^2 + \frac{1}{\left(1
 - \frac{2M}{r} + \frac{Q^2}{r^2}\right)}dr^2 + r^2\left(d\theta^2 +
 \sin^2\theta d\phi^2 \right)
\label{eq1}
\end{equation}
We first consider the embedding of the $(r-t)$ plane in $(2+1)$-Minkowski
spacetime. Following \cite{dm}, let us reduce the metric to the $(r-t)$ plane,
\begin{equation}
ds^2 = -\left(1 - \frac{2M}{r} + \frac{Q^2}{r^2} \right)dt^2 + \frac{1}{\left(1
 - \frac{2M}{r} + \frac{Q^2}{r^2}\right)}dr^2   \label{eq2}
\end{equation}
which we wish to embed into the flat $(T,X,Y)$ spacetime. There are two 
horizons located at
\begin{equation}
r_{+} =  M + \sqrt {M^2 - Q^2}\ ,\ \ \,
r_{-} =  M - \sqrt {M^2 - Q^2}  \label{eq3}
\end{equation}

By considering the intermediate coordinates
\begin{equation}
\rho = \sqrt {X^2 - T^2}\ ,\ \ \,
\psi = \tanh^{-1} \left(T/X\right)  \label{eq4}
\end{equation}

$ \ $   in the region $r > r_{+}$
                and
\begin{equation}
\rho = \sqrt {T^2 - X^2}\ ,\ \ \,
\psi = \tanh^{-1} \left(X/T\right)  \label{eq5}
\end{equation}
        $ \  $        in the region $r_{-} < r < r_{+}$,
it is possible to show that the embedding is completely specified by taking
\begin{equation}
Y\left(r\right) = \int_{r_{+}}^r \sqrt {\frac{1 - \frac{1}{4\kappa^2}
 \left(\frac{d\mu}{dr}\right)^2}{\mu}} dr    \label{eq6}
\end{equation}
         where
\begin{equation}
\mu = 1 + 2\Phi = 1 - \frac{2M}{r} + \frac{Q^2}{r^2} \ ,\ \ \,
\kappa = \frac{1}{2}\frac{d\mu}{dr} |_{r = r_{+}}. \nonumber
\end{equation}
As shown in \cite{dm}, note that the radical in the integrand near the horizon
would go as $-\frac{d^2\mu/dr^2}{2\kappa^2}$ which implies
that embedding cannot proceed when $\frac{d^2\mu}{dr^2}$ turns
positive. That would happen at $r = 3Q^2/2M$ which would lie
between the two horizon unless $Q^2/M^2 > 8/9$. Fig. 1 shows the
embedding diagram. The two protruding flanges indicate the two
asymptotically flat regions while the two half cones represent the
corresponding region enclosed by the two horizons.

\begin{figure}[tp]
\includegraphics[width=7cm]{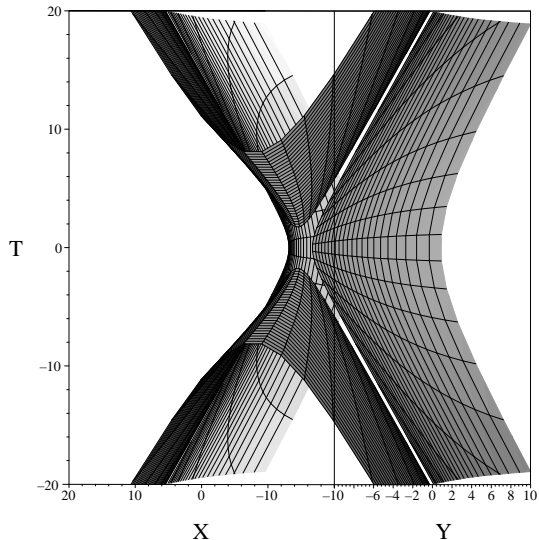}
\caption{Embedding diagram for RN spacetime
($r > r_{+}$). The flanges represent the asymptotically flat
regions. The cones represent their respective interiors. }
 \label{fig1}
\end{figure}

The scalar curvature of the $(r-t)$ plane is given by
\begin{equation}
R = \frac{2\left(2Mr - 3Q^2\right)}{r^4} \label{eq7}
\end{equation}
That is embedding can proceed only until $R=0$ and not beyond it.

 We now turn to the embedding of a constant time slice, the $(r-\phi)$ plane
of RN geometry. Following \cite{dm} as before, we have to consider
embeddings both in the Euclidean as well as Minkowski flat space depending
upon the signature of the $(r-\phi)$ plane. The constant time 
equatorial plane slice would have the reduced metric,
\begin{equation}
ds^2 = \frac{1}{1 - \frac{2M}{r} + \frac{Q^2}{r^2}} dr^2 + r^2 d\phi^2
\label{eq8}
\end{equation}

For $r > r_{+}$,$ \ r$ it is spacelike and we embed this metric into the 
Euclidean
 space. To embed it, we employ the cylindrical polar coordinates in the
 Euclidean space, with the angle $\phi$ being common for both. Thus
the Euclidean metric is
\begin{equation}
ds^2 = d\rho^2 + \rho^2 d\phi^2 + dz^2 \label{eq9}
\end{equation}
For constant $z$, a translation in $\phi$, $ \ \left(r,\ \phi\right)\rightarrow
 \left(r, \ \phi + \delta\phi\right) \ $ and $ \ \left(\rho, \ \phi\right)
\rightarrow \left(\rho, \ \phi + \delta\phi\right) \ $ must produce the same
 proper displacement in both metrics.
Hence we must set
\begin{equation}
\rho = r \label{eqn10}
\end{equation}
Now the embedding can be completed by specifying $z$ as a function of $r$. This
is done by requiring agreement of the metrics (8) and (9) at
constant $\phi$. Thus
\begin{equation}
d\rho^2 + dz^2 = \frac{1}{1 - \frac{2M}{r} + \frac{Q^2}{r^2}} dr^2 ,
 \nonumber
\end{equation}

which can be solved to give (setting $\rho = r$ )
\begin{equation}
z\left(r\right) = \int_{r_{+}}^r \sqrt{\frac{2Mr - Q^2}{r^2 - 2Mr + Q^2}} dr
\label{eqn11}
\end{equation}
where we have set $z\left(r_{+}\right) = 0 \ $ and have chosen the appropriate
 sign for $z$. That means there is no problem until $r = Q^2/2M < r_{-}$. Here
 we have to consider all the three patches for embedding. For $r > r_{+}$, 
the embedding diagram is shown in Fig. 2, which is quite similar to the 
Schwarzschild (see \cite{n1}).

\begin{figure}
\includegraphics[width=7cm]{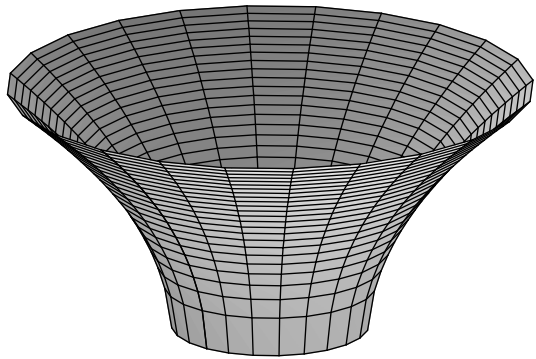}
\caption{Embedding of $(r-\phi)$ plane of RN spacetime for $r > r_{+}$ .
 One asymptotically flat region is shown. The lower end of the 'funnel'
 corresponds to the outer horizon at $r_{+}$.} \label{fig2}
\end{figure}

 $ \ $
 For $r_{-} < r < r_{+} $, $ \ \left(1 - 2M/r + Q^2/r^2\right) < 0 $ and the
 coordinate $r$ is timelike. We therefore embed the metric (8),
rewritten for convenience as
\begin{equation}
ds^2 = \frac{-1}{\frac{2M}{r} - 1 - \frac{Q^2}{r^2}} dr^2 + r^2 d\phi^2
\label{eqn12}
\end{equation}
into $(2+1)$-Minkowski spacetime given by the metric,
\begin{equation}
ds^2 = -dz^2 + d\rho^2 + \rho^2 d\phi^2.  \label{eqn13}
\end{equation}
Once again we set $\rho = r \ $. This reflects the fact that though $r$ is
 timelike, it still functions as the area radius of the spacetime. Proceeding
as before, we require that
\begin{equation}
-dz^2 + d\rho^2 = \frac{-1}{\frac{2M}{r} - 1 - \frac{Q^2}{r^2}} dr^2
\nonumber
\end{equation}
from which we get
\begin{equation}
z\left(r\right) = \int_{r}^{r_{+}} \sqrt{\frac{2Mr - Q^2}{2Mr - r^2 - Q^2}} dr
\label{eq14}
\end{equation}
where we have again set $z\left(r_{+}\right) = 0 \ $.
The embedding diagram is shown in Fig. 3. The top of Fig. 3
would join smoothly to the bottom of the funnel of Fig. 2.

\begin{figure}[bp]
\includegraphics[width=7cm]{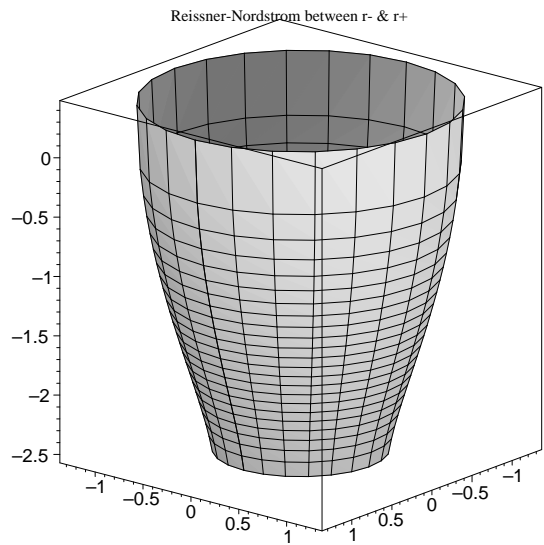}
\caption{Embedding diagram for the $(r-\phi)$ plane of RN spacetime for
$r_{-} < r < r_{+} \ $. On the same scale as Fig. 2, the top of this diagram
would join smoothly to the bottom of the one in Fig. 2.}
\label{fig3}
\end{figure}

                                                                                   $ \ $
For  $r < r_{-} \ $, $r$ is once again spacelike and the embedding is carried
 out as before (Eqn. (8) through Eqn. (11)) , except that the limits of
integration in
 the expression for $z$ would now range from $r$ to $r_{-}$. This time,
 however, the expression under the radical in eqn. (11) changes sign at 
$r = 3Q^2/2M \ $ and that is where the embedding terminates.
The diagram (blown up to see the detail) is shown in Fig. 4.
Again, on the same scale, the diagrams in Figs 2 and 3 would
smoothly join at the ends.

\begin{figure}[tp]
\includegraphics[width=6cm]{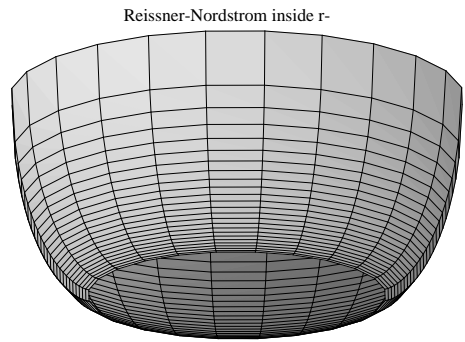}
\caption{Embedding of the $(r-\phi)$ plane of RN spacetime for $r < r_{-} \ $
. The diagram is enlarged in order to see the detail.}
\label{fig4}
\end{figure}

The scalar curvature for this metric is
\begin{equation}
R = \frac{-2\left(Mr - Q^2\right)}{r^4} \label{eq15}
\end{equation}
which changes sign at $r = Q^2/M$, lying between the two
horizons. Unlike the $(r-t)$ plane case, where embedding
terminated when curvature turned zero, here the embedding can
proceed beyond $r = Q^2/M$ as far in as $r = Q^2/2M$. In this case
we have embeddings for both positive as well as negative curvature
regions. It begins with negative curvature in Fig. 2, changing sign in Fig. 3,
and it is all positive in Fig. 4.

Fig. 1 refers to the whole field incorporating the contributions of both
potential and field energy while the spatial embedding diagrams as shown in
 Figs 2-4 refer to the contribution of the field energy alone. The
embedding diagrams do reflect when the curvature changes sign from
negative to positive. This would mean the contribution of field
energy changing its sense {\it from attractive to repulsive} and
it happens at $r = Q^2/M$, where interestingly $d\Phi/dr$ also
changes sign;i.e. gravitational field changes sign \cite{n3}. As mentioned 
earlier, the contributions from the
potential as well as the field energy act in unison and hence they
should change their sense at the same radius. The point is that for 
attractive gravity, the field energy must be negative and the vice-versa 
\cite{n2}. That is why the two have to change sign at the same radius.
 
 However, the difference is that the field energy only links to the spatial
curvature \cite{n0} while the other to the potential gradient. This happens 
because curvature in the $(r-\phi)$ plane is proportional to $d\Phi/dr$ while  
the curvature in the $(r-t)$ plane 
measures the tidal acceleration between two neighbouring geodesics which is 
given by $d^2\Phi/dr^2$, and the embedding in this case terminates where it 
changes sign at $r = 3Q^2/2M$. On the other hand the embedding for the 
$(r-\phi)$ case terminates where the gravitational potential $\Phi$ changes 
sign at $r = Q^2/2M$. It is interesting that 
embedding of the $(r-t)$ plane (i.e. the whole field) hinges on the tidal 
acceleration while that of the $(r-\phi)$ plane on the potential. 
This feature we do not quite understand and would like to probe 
it further in future.

 $ \ $

ACKNOWLEDGEMENTS: We would like to thank Donald Marolf for reading the 
first draft of the manuscript. AP would like to thank Jawaharlal Nehru Centre
for Advanced Scientific Research for the summer fellowship which
facilitated this work and IUCAA for hospitality.

\end{document}